\documentstyle[12pt]{article}
\begin{document}
\begin{titlepage}
\vspace{2.5cm}
\baselineskip 16pt
\rightline{hep-ph/9609232 ~~(Version 3, July 1998)}
\begin{center}
\large\bf{Could Spontaneous Transitions be Spontaneous?$^1$}\\
\vspace{1.5cm}
\large{
Bernd A. Berg$^{2,3}$}\\
\end{center}
\vspace{3cm}
\begin{center}
{\bf Abstract}\\
\end{center}

It is considered to re-formulate quantum theory as it appears: A
theory of continuous and causal time evolution, interrupted by 
discontinuous and stochastic jumps.
Relying on a global reduction process, a hypothesis is introduced
postulating spontaneous collapse of superpositions
of states which describe spontaneous absorption or emission. The
collapse probability determines a mean collapse time 
$\tau^c = b \hbar / \triangle E$, where $b$ is a dimensionless 
constant and $\triangle E$ is the difference in energy distribution
between alternative
branches. Ramsey atomic beam spectroscopy yields a lower bound on $b$ 
and avalanche photodiodes give an upper bound, such that 
$1.35\cdot 10^{11} < b < 3.8\cdot 10^{21}$.
\vfill

\footnotetext[1]{{Work partially supported by the Department of 
                  Energy under contract DE-FG05-87ER40319.}}
\footnotetext[2]{{Department of Physics, The Florida State University,
                      Tallahassee, FL~32306, USA. }}
\footnotetext[3]{{Supercomputer Computations Research Institute,
                      Tallahassee, FL~32306, USA.}}
\footnotetext[4]{{E-mail: berg@hep.fsu.edu; Fax: (850) 644 6735.}}

\end{titlepage}

\baselineskip 16pt


A photographic plate consists of an emulsion which separates grains
of AgCl (or similar) molecules. Let me present a simplified 
discussion of the detection of a single photon. The photon may 
dissociate an AgCl molecule. The Cl
escapes, whereas the Ag radical starts, upon photographic 
development, a chemical reaction which
leaves a small, visible spot. Before detection the 
one-particle photon wave may pass through {\it both} parts of a 
double slit, see for instance ref.\cite{Bra}. 
The geometry of the device (distance between the
slits etc.) can be chosen such that the photon is spread out over
a region covering distances much larger than a single grain, which
determines the size of the finally visible spot. The physics which
causes the photon wave to collapse is not understood.

Let us first recall where the Schr\"odinger equation 
leaves us. For simplicity we assume that each grain consists of
precisely one AgCl molecule. We label the AgCl molecules on
the plate by $i=1,...,n$ and assume that the relevant features
of the photographic plate are described by products of
AgCl molecule wave functions. Initially the state is 
\begin{equation}
|\Psi\rangle = |\Psi_0\rangle = \prod_{i=1}^n |\psi_i^b\rangle ,
\ \langle \Psi_0 | \Psi_0 \rangle = 1, \label{Psi0}
\end{equation}
where the 
$|\psi_i^b\rangle$, $(i=1,...,n)$ indicate the {\it bound}
AgCl molecules and the overlap between different molecules is
neglected, {\it i.e.} $\langle \psi_i^b | \psi_j^b \rangle =
\delta_{ij}$. Through interaction with the photon the state
is transformed into
\begin{equation} \label{causal}
|\Psi\rangle = c_0 |\Psi_0\rangle + \sum_{j=1}^n c_j 
|\Psi_j\rangle ~~{\rm with}~~ |\Psi_j\rangle =
|\psi_j^d\rangle \prod_{i\ne j} |\psi_i^b\rangle ,\
\langle \Psi_j | \Psi_j \rangle = 1\, . \label{Psi}
\end{equation}
Here $|\psi_j^d\rangle$, $(j=1,...,n)$ denotes a {\it dissociated}
AgCl molecule. A quantum measurement is constituted by
the fact that only one of the $|\Psi_j\rangle$, $(j=0,1,...,n)$
states survives, each with probability $P_j=|c_j|^2$,
$\sum_{j=0}^n P_j = 1$. The probabilities $P_j$ are related to
the photon wave function $\psi_{ph}$ by means of 
\begin{equation}
 P_j = const\ \int_{V_j} d^3x\ |\psi_{ph}|^2\, ,\
j=1,...,n\, . \label{probability}
\end{equation}
The constant does not depend on $j$, and $V_j$ is a
cross-sectional volume corresponding to the $j$th molecule.
Picking a branch $|\Psi_j\rangle$ becomes in this way
interpreted as observing the photon at the position $V_j$
(for a perfect
detector $c_0=0$). It is remarkable that in this process
of measurement the photon becomes destroyed through 
spontaneous absorption by the dissociating molecule. To
summarize, measurements perform wave function {\it reductions} by
making decisions between alternatives proposed by the continuous, causal
time evolution part of Quantum Theory (QT). In our example
the reduction decides the location of the visible spot. Given the photon
wave function $\psi_{ph}$, QT predicts probabilities for the reduction
alternatives.

Whereas time evolution from eqn.(\ref{Psi0}) to
(\ref{Psi}) is described by the Schr\"odinger equation,
this is not true (or at least controversial) for the
measurement process. Decoherence theory, for
an overview see \cite{Zurek}, tries to establish that
conventional time evolution leads, in the situation of
eqn.(\ref{Psi}), to $\langle\Psi_j |
\Psi_k\rangle = 0$ for $j\ne k$, such that it becomes
impossible to observe contradictions with interference
effects predicted by the Schr\"odinger equation. In
contrast to this, explicit collapse models predict
deviations from conventional QT, see
for instance \cite{Ch86}. So far, no such deviations
have been measured.
The Schr\"odinger equation, more precisely its applicable 
relativistic generalization $|\Psi (t)\rangle = \exp (
- H\, t)\, |\Psi (0)\rangle $, describes a continuous and
causal time evolution and I shall use the notation Quantum 
Object (QO) for matter $|\Psi (t)\rangle$ as long as it 
exhibits this behavior. Concerning the measurement process,
it seems to be widely believed that many body 
processes, involving $\gg 10^{10}$ particles \cite{Bra}, are
responsible. In the presented example the photographic plate, 
possibly also the environment beyond, would be blamed. However, 
the ultimate collapse into one branch $|\Psi_j\rangle$ remains 
an unexplained property. No satisfactory derivation from the known
properties of microscopic matter appears possible. Consequently, a
search for hereto overlooked new, fundamental properties of matter
is legitimate.

The question arose whether hidden variables may exist which
ensure local, continuous and causal time development for the
entire system, including measurements. Bell \cite{Bell} turned
this apparently philosophical question into physics by showing
that all such local, realistic theories are measurably distinct
from QT (Bell's inequalities). Subsequently, many experiments 
were performed and local, realistic theories are now 
convincingly excluded. For instance, the experiments of \cite{Aspect}
found violations of Bell's inequalities for spacelike measurements
on entangled quantum states. In such experiments one performs 
measurements at distinct locations, say $\vec{x}_1$ and $\vec{x}_2$, 
in time intervals small enough that any mutual influence through 
communication at 
or below the speed of light can be excluded. Results at $\vec{x}_1$
correlate with those at $\vec{x}_2$ (and vice versa) in a way that
{\it excludes} an interpretation as a classical correlation, 
see \cite{Mermin} for a pedagogical presentation. Such 
effects underline the need for qualitatively new properties of matter, 
because they cannot be propagated through local, relativistic wave
equations (which also govern the interaction with the environment).

Also a consistent description of the space-time evolution of the quantum
state vector $|\Psi\rangle$ under such measurements encounters 
difficulties. After accepting that a measurement at $(c\, t_1,\vec{x}_1)$ 
or $(c\, t_2,\vec{x}_2)$ interrupts the continuous, causal time evolution
by a discontinuous jump, once faces the problem that Lorentz 
transformations can change the time ordering of spacelike events. In
a recent paper \cite{Be98a} it has been shown that a spacetime picture
for a physical state vector with relativistically covariant reduction
exists. It may be summarized as follows:

\begin{description}
\item{(1)} Measurement are performed by detectors, which are part of
the state vector, at localized spacetime positions
$(c\, t_i,\vec{x}_i)$, $i=1,2,\dots\,$.
\item{(2)} Discontinuous reductions of the state vector are defined
on certain Lorentz covariant spacelike hypersurfaces, which in some
neighbourhood of a detector include its backward light cone.
\item{(3)} The thus defined measurements happen in some reduction order,
which is {\it not} a time ordering with respect to a particular inertial
frame.
\end{description}

Based on this scenario, I pursue in the present paper an approach which 
builds on the strengths of QT and tries to supplement it with new laws
for reductions, such that the conventional rules 
for measurements (Born's probability interpretation) follow. These laws 
are supposed to act on the {\it microscopic} level, independently of 
whether macroscopic measurements are 
actually carried out or not. Typically, they will effect some
interference phenomena. This implies observable consequences and
makes their eventual existence a physical issue. On the other hand,
we will see that rules can be designed in a way that most interference
effects survive entirely, whereas those affected are only weakened in
the sense of a decreased signal over background ratio (visibility). That
makes such laws difficult to detect, as most experiments work with
ensembles of particles and are happy to demonstrate a small signal over
a large (subtracted) background. Fortunately, recent years have seen
considerable improvements of experimental techniques, such that invoking 
experimental input may become feasible.

The central idea of my approach is to {\it propose} that the Ability 
To Perform Reductions (ATPR) between alternatives proposed by QT is a
hereto unidentified {\it elementary} property of microscopic matter. 
I shall use the notation Quantum Detector (QD) to denote microscopic as
well as macroscopic matter acting in its ATPR. Matter gets such 
a dual character: As QO it follows the continuous, deterministic time 
evolution. As QD it has the ATPR and causes jumps in the wave function.
The goal is to explain the functioning of actually existing
macroscopic detectors from the properties of microscopic QDs.
Within our hypothetical framework central questions are now:
\begin{description}
\item{(1)} Which conglomerates of matter constitute a QD?
\item{(2)} Which are precisely the alternatives of QT standing
           up for reductions?
\item{(3)} What are the rules according to which QDs make their
           reductions?
\end{description}

It is unlikely, that ultimate answers can be found without
additional experimental guidance. But it is instructive to introduce
a simple hypothesis which allows (a) to illustrate the possibilities and
general direction of the approach and (b) focuses on questions about
experimental input which, quite generally, may be crucial for 
achieving progress in the field. 

Let us return to the detection of a photon by a photographic plate.
The simplest possibility is to attribute to each {\it single} AgCl 
molecule the ATPR about collapsing the photon wave function. As this 
is a spontaneous absorption, we get to the question asked in
the title of this paper. We assume that each molecule acts 
independently when making its reductions and, by chance,
the $j^{th}$ molecule makes its reduction first, ahead of the 
others. The alternative is to decay or to stay intact. Either choice
causes a jump in the wave function $|\Psi\rangle$ of eqn.(\ref{causal}). 
Subsequently rules are given which seem (a) to be minimal and (b)
consistent with observations.

The collapse results are fixed by the rules of quantum mechanics.
Namely, to be either ($f$ stands for final)
\begin{equation} \label{collapse1}
|\Psi\rangle \to |\Psi\rangle_f = {c_j\over \sqrt{P_j}}\, 
|\Psi_j\rangle ~~~{\rm with\ probability}~~ P_j = |c_j|^2 
\end{equation}
or
\begin{equation} \label{collapse2}
|\Psi\rangle \to |\Psi'\rangle = \sum_{k\ne j} c'_k\, |\Psi_k\rangle ,\ 
c'_k= {c_k \over \sqrt{1-P_j}}, ~~~{\rm with\ probability}~~ 1-P_j\, . 
\end{equation}
The $k$-sum in eqn.(\ref{collapse2}) includes $k=0$, compare 
(\ref{causal}). The particular choice of the phase factors,
$c_j/\sqrt{P_j}$ and $c_k/\sqrt{1-P_j}$, assumes that a decoherence
process leads into alternative branches, whereas for collapse with
incomplete decoherence the issue would have to be resolved by the
collapse rules. Each molecule thus constitutes a QD. As their mutual
distances are short and their
relative motion is negligible, compared to the speed of light, we can
ignore the relativistic complications discussed in \cite{Be98a} (the
backward light-cone becomes an excellent approximation to 
``instantaneous'').

Equation (\ref{collapse1}) 
implies as final result a dark spot at the position of molecule $j$. 
By construction this happens with the correct probability $P_j$. As soon 
as the wave function (\ref{collapse1}) rules, the reduction is
completed. This is different when molecule $j$ 
stays intact. Then the same rules (\ref{collapse1}) and 
(\ref{collapse2}), which collapse $|\Psi\rangle$ of eqn.(\ref{causal}), 
have now to be applied to the wave function $|\Psi'\rangle$ of 
eqn.(\ref{collapse2}). Assume, molecule $l$ (note $l\ne j$ as the 
branch $|\Psi_j\rangle$ does no longer exist) makes the next reduction. 
The transformation will be either
\begin{equation} \label{collapse3}
|\Psi'\rangle \to |\Psi\rangle_f = {c'_l\over \sqrt{P'_l}}\, 
|\Psi_l\rangle ~~~{\rm with\ probability}~~ P'_l=|c'_l|^2
\end{equation}
or
$$ |\Psi'\rangle \to |\Psi''\rangle = \sum_{k\ne j,l} c''_k\, 
|\Psi_k\rangle ,\ c''_k= {c'_k \over \sqrt{1-P'_l}}, 
~~{\rm with\ probability}~~ 1-P'_l\, . $$
Putting equations (\ref{collapse2}) and (\ref{collapse3}) together, 
we obtain
$$ |\Psi\rangle \to |\Psi\rangle_f = {c_l\over \sqrt{P_l}}\, 
|\Psi_l\rangle ~~~{\rm with\ probability}~~ P_l,$$
{\it i.e.} precisely the correct likelihood to find the dark
spot at the position of molecule $l$. Continuing the procedure,
it is easy to see that all probabilities come out right.

Once a molecule $j$, $j=1,...,n$ has collapsed $|\Psi\rangle$
into the $|\Psi\rangle_f$ state of eqn.(4), the chemical reaction$^1$
\footnotetext[1]{{During the chemical reaction similar 
collapse processes may continue. Presently, they are not of 
interest to us, as our aim is to discuss the collapse of the 
incoming photon wave function, which has the special property of 
being transversally spread out over a macroscopic region.}} 
-- initiated by the corresponding branch of each molecule -- survives 
only in the neighbourhood of molecule $j$, where the visible spot 
will occur. Let $t=0$ be the time at which the photon hits the 
photographic plate. This time is well-defined as long as we can 
assume that the photon flight time over a distance of the relevant 
thickness of the photographic plate (for example $0.3~mm \Rightarrow 
\triangle t = 10^{-12}~s = 1~ps$) is much smaller than the typical 
collapse time. Let us denote by
\begin{equation} \label{tauf}
P^f(t)=(1-P_0)\, (1-e^{-t/\tau^f(t)})
\end{equation}
the probability that, at time $t$, the (entire) system has 
decided about the location of the dark spot. Here $P_0=|c_0|^2$, see 
eqn.(\ref{causal}), is the probability that the system fails to 
detect. The r.h.s. of (\ref{tauf}) defines the system collapse time 
$\tau^f(t)$. If $\tau^f$ is constant, it is the mean time the system 
needs to make its reduction (with corresponding collapse probability 
density $(\tau^f)^{-1}\, \exp (-t/\tau^f)$).

Let us assume that each molecule performs reductions on its own and that
$\rho^c_j(t)$ is the likelihood per time unit that the $j^{th}$ molecule
makes its reduction. We simplify the situation further
and consider a $\rho^c_j(t)$ that does not 
depend on $j$ and is a step function: $\rho_j^c(t)=\rho^c\, \theta(t)$
with $\rho^c =$ constant. The corresponding one molecule collapse 
probability is
\begin{equation} \label{tauc}
p^c(t)=(1-e^{-t/\tau^c})\, \theta(t)\, ,
\end{equation}
where $\tau^c=1/\rho^c$ is the mean collapse time of a single molecule. 
The molecules make their reductions in some sequential order. For our 
purposes the reductions process comes to a halt as soon as one molecule
has decayed. Assume, $n^c(t)$ molecules made their reductions. Whatever
values the $P_j$ in eqn.(\ref{collapse1}) take,
$P^f(t) = (n^c(t)/ n)\, (1-P_0)$ is the probability that the collapse 
process has selected a definite location. As $n^c(t)=n\, p^c(t)$, we 
conclude
\begin{equation}
P^f(t)= (1-P_0)\, p^c(t)\, .
\end{equation}
With the approximations made the system collapse time $\tau^f$, defined
by (\ref{tauf}), and the single molecule collapse time $\tau^c$, 
defined by (\ref{tauc}), are identical. 
Soon some arguments will be given that the constant $\tau^c$
should be regarded as upper bound of the system collapse time 
$\tau^f(t)$. 

Let us return to the central questions. In our discussion of 
detection of a spread-out photon, I assumed the following:
(1) Each, single AgCl molecule may act as QD. (2) One alternative
stands up for reduction: decaying (through absorbing the photon)
or staying intact. (3) Each AgCl makes its reduction with a certain,
constant likelihood per time unit: $\rho^c$.

Ad (1): Assuming that a single AgCl molecule can act as QD reflects
the attempt to introduce an ATPR as a fundamental property of
microscopic matter. In our simplified discussion each AgCl molecule
is separated from the others by the emulsion and causal interactions
between them can be neglected. In a real photographic film only grains
of AgCl molecules are separated. Causal interactions between an AgCl 
molecule and its  neighbors within one grain cannot be neglected. 
Indeed, the initiated chemical process will spread out over the
entire grain. If the collapse time is sufficiently large, competing
(ultimately alternative) chemical processes would start to evolve in 
several grains. Under such circumstances, the definition of the QD
should be extended to include each causally connected region of AgCl
molecules. As a general rule, I find
it attractive to conjecture that a  conglomerate of matter which (in a
reasonable approximation) can be treated as isolated QO can also be
regarded as isolated QD. Neglecting the influence of most of the world 
is precisely how we get solutions out of QT. The hypothesis is, whenever
this works well for a QO, this QO may also constitute a QD whose
reduction probabilities are determined by its local quantum state,
although this quantum state may participate in discontinuous, non-local 
transformations.

Ad (2): The scenario, pursued now, is that the QT alternatives up 
for reductions have, quite generally, to do with absorption and emission 
of particles. Here I limit the discussion to the absorption and emission 
of photons, the process argued to be at the heart of every real, existing 
and functioning measurement device. Of course, other physical processes 
(like for instance in nuclear decay) should then be governed by similar
rules. In essence: Ruled by not yet identified {\it stochastic} laws,
superpositions of Fock space sectors with distinct particle numbers are 
conjectured to collapse into particle number eigenstate sectors.

Ad (3): Our ATPR introduces an explicit arrow in time. This is attractive, 
because it is a matter of fact that such an arrow exists. The canonical 
conjugate variable to time is energy. Therefore, a frequency law which
relates the collapse time to the difference $\triangle E$ in energy
distribution between emerging branches is suggested
\begin{equation} \label{tcguess}
\tau^c\ =\ \tau^c (\triangle E)\, ,
\end{equation}
where the energy difference is defined as the one experienced by the
QD. For example, in case of our single AgCl molecule the difference
between absorbing or not absorbing the photon is: 
$\triangle E = E_{\gamma}$, where $E_{\gamma}$ is the energy
of the photon. The total energy is (in the same way as in QT) conserved 
in our approach.

The introduced system has been chosen because of its popularity in QT 
text books in connection with the double slit effect. 
Instead of the photographic plate other measurement devices
can be considered. For instance, the emergence of a track in a
bubble chamber through ionization by an high energy particle allows a 
similar discussion. Here it is instructive to consider the emergence of 
such a track in monatomic dilute gas, say hydrogen. Assume that {\it one}
incoming high energy particle has been split into two distinct 
transversally sharp rays$^1$, 
\footnotetext[1]{{Within our approach it might, however, happen that such 
a state collapses spontaneously, because the device which caused the
split (and hence correlates with it) might act as a QD.}}
each with 50\% probability content. At time $t=0$ the two rays may hit 
spacelike regions of 
hydrogen gas. Each ray builds up a column of half-ionized atoms. Let us 
focus on one of them, consisting of $n$ participating atoms. If one of 
the atoms of our column emits a photon by re-capturing an 
electron the relevant reduction has been made. A transformation of 
type (\ref{collapse1}) puts all atoms of
the competing column into their unperturbed branches and the atoms
of our column into their ionized branches. There is now some ambiguity
about what should be considered a QD. Should each single atom
(including the involved electrons) act as independent QD or should all
atoms of the column together form one, single QD? In favor of the first
viewpoint is that the gas is assumed to be dilute. Hence, the mutual
influence through continuous, causal time evolution between the atoms is
negligible. On the other hand, the high energy particle correlates all
the atoms within the column (and, of course, also the other column): If
one atom performs its reduction in favor of the ionized branch, all
other are put there too. Assume the atoms act independently and the
differences in energy distribution between their branches are
$\triangle E_i$, ($i=1,...,n$), implying corresponding mean collapse
times $\tau^c_i (\triangle E_i)$. The probability that none
of them makes the reduction during the time interval $[0,t]$ becomes
$$ q^c (t) = \prod_{i=1}^n \exp [-t/\tau^c_i (\triangle E_i) ] 
 = \exp \left[ - \sum_{i=1}^n t/\tau^c_i (\triangle E_i) \right]\, .$$
On the other hand, if they all together form one single QD, this
probability becomes 
$$ q^c (t) = \exp [-t/\tau^c(\triangle E)]\, ,~~{\rm where}~~
\triangle E = \sum_{i=1}^n \triangle E_i $$
is the total difference in energy distribution between the alternative,
macroscopic branches. Remarkably, the $q^c(t)$ of the last two equations
agree, when the law for the mean collapse time is
\begin{equation} \label{berg}
 \tau^c (\triangle E)\ =\ {b\, \hbar \over \triangle E}\, ,
\end{equation}
where $b$ is a dimensionless constant. (Correspondingly, $\tau^c_i
=b\, \hbar/\triangle E_i$, $i=1,..,n$, of course.) Equation (\ref{berg})
has phenomenologically attractive features. The first one is that the 
indicated ambiguity is rendered irrelevant. Another is that the
collapse time becomes large for small energy differences. Especially,
superpositions of states degenerate in energy will not collapse.
In this context {\it a measurement device is now an apparatus which 
speeds up the collapse by increasing the difference in energy
distribution between quantum branches.} Before the distinct branches
become macroscopically visible, the energy difference becomes so large
that collapse happens with (practical) certainty.

Are there observable consequences beyond standard QT?
Reduction by an AgCl molecule destroys the possibility of
interference of the branches (\ref{collapse1}) and (\ref{collapse2}) 
of the wave function (\ref{Psi}). In particular, this does still hold 
for the case of a single molecule $(n=1)$. But it appears unlikely that
anyone will, in the near future, measure interference effects between 
AgCl$+\gamma$ and Ag$+$Cl. Hence, there is no contradiction. 
In addition, it should be noted that our mechanism leaves the most
commonly observed interference effects intact: Namely, all those which
rely on the wave character of particles in a Fock space sector with
fixed particle number. This includes photon or other particle waves 
passing through double slits and so on. Neutron interferometry which
relies on hyperfine level splitting would, in principle, be 
suppressed. However, the energy differences are small such that 
observable effect are unlikely. 

Larger energy differences are achieved in atomic beam spectroscopy. 
Ramsey fringes have been observed from interference of branches which 
differ by photon quanta with energy in the $eV$ range. Figure~1 depicts
the interaction geometry of Bord\'e \cite{Bo84}, for a recent review
see~\cite{St97}. An atomic beam of two level
systems $(E_0 < E_1)$ interacts with two counterpropagating sets
of a traveling laser wave. The laser frequency is tuned 
to the energy difference $\triangle E=E_1-E_0$, such that
induced absorption/emission processes take place at each of the
four interaction zones. The laser intensity is adjusted such that at 
each interaction zone an incoming partial wave is (further) split 
into two equally strong parts, $|\psi_0,m_0\rangle$ and 
$|\psi_1,m_1\rangle$. Here $|\psi_0\rangle$ denotes an atom in its 
incoming state, $|\psi_1\rangle$ an excited atom 
and $m_0$, $m_1$ are the numbers of photon moments transferred. 
Examples are indicated in the figure. The process leaves us with $2^n$ 
partial waves after the $n^{th}$ interaction zone, $n=1,2,3,4$. Of the 
final sixteen partial waves $4\cdot 2=8$ interfere under detuning of the 
laser frequency. Positions and directions of those eight partial waves 
are along the four lines, indicated after the last interaction 
zone of figure~1. The interference can be made visible by monitoring
the decay luminosity $I$ of the excited states $|\psi_1\rangle$ after 
the last interaction zone. The contrast or visibility is defined by
\begin{equation} \label{contrast}
K = {I_{\max} - I_{\min} \over I_{\max} + I_{min}}\, ,
\end{equation}  
where $I_{\max}$ and $I_{\min}$ are maximum and minimum of the measured
luminosities. Eight of the final sixteen partial waves are in excited
states and four of them interfere in two pairs, $|\psi_1,-1\rangle$
and $|\psi_1,1\rangle$ of the figure. Hence, the optimal contrast for the 
Bord\'e geometry is
\begin{equation} \label{Kopt}
K_{opt} = {(4-4) + (8-0)  \over (4+4) + (8+0)} = 0.5\, .
\end{equation}
This result is found by normalizing (in arbitrary units) the average
luminosity of each excited partial wave to one. Four decoherent branches
contribute then $I_{\max}=I_{\min}=4$, whereas the other four excited
partial waves contribute $I_{\min}=0$ and $I_{\max}=8$ (for $I_{\min}$
they annihilate one another and in the other extreme they amplify). 

According to our hypothesis, integer photon numbers get restored with
a collapse time $\tau^c=\tau^c(\triangle E)$. If this happens in 
range~1 of figure~1, the interference effect becomes 
entirely destroyed. The likelihood for it to happen is $p^c=1-q^c$, 
where $q^{c}=\exp (-t_D/\tau^c)$ and $t_D$ is the time an atom stays 
in range~1. In range~2 each of the two $|\psi_1,1\rangle$ partial waves
has borrowed $1/4$ of a photon from the laser beam. To get a unique
collapse description, we invoke a minimality assumption:
The splitting of the atom
has to be constructed with the minimal number of photons possible.
The assumption seems to be natural, because two photons with
the same quantum numbers cannot be distinguished. It follows that
the system can collapse either into the two $|\psi_1,1\rangle$
partial waves or into the two $|\psi_0,0\rangle$ partial waves. 
Neither collapse has observable consequences,
because the interference effects of the upper part and lower
part of figure~1 are not distinguished by measuring the decay 
luminosity. In range~3 two split photons (distinct momenta) get 
involved: One mediates 
collapse between $|\psi_1,1\rangle$ and $|\psi_0,2\rangle$, the other 
between $|\psi_0,0\rangle$ and $\psi_1,1\rangle$. These two collapse 
processes are supposed to act independently. Each destroys, if it happens, 
half of the interference effect. The probability for both of them to 
happen is $(p^c)^2$ (using that $t_D$ is identical in range~3 and~1) and 
the probability that one of them (excluding both) happens is 
$p'^c=1-(q^c)^2-(p^c)^2$. Putting things together, the optimal contrast 
becomes
\begin{equation} \label{Kcopt}
K_{opt}^c=16^{-1}[8-8\,p^c-4\,q^c\,p'^c-8\,q^c\,(p^c)^2]=
0.5\, \exp [ -2\, t_D/\tau^c (\triangle E)]\, .
\end{equation}
Experiments performed at the Physikalisch-Technische Bundesanstalt
(PTB) Braunschweig rely on the $^3P_1$--$^1S_0$ transition of $^{40}Ca$ 
which has $\lambda = 657.46$~nm, {\it i.e.} $\triangle E =1.886\, 
eV$. The best contrast achieved \cite{Riehle} is approximately
$K=0.2$ with $t_D=21.6\cdot 10^{-6}\,s$. The actual experiments
are performed using pulsed laser beams applied to laser cooled 
atoms in a magneto-optical trap. The times $t_D$ and $t_d$,
corresponding to the distances $D$ and $d$ of figure~1, are 
then the times between the laser pulses, see~\cite{St97} for
details. Relying on the PTB result we obtain the estimate
$$ \tau^c_{\min}\, (1.89\, eV) = 2\,t_D\, /\, \ln (5/2) = 
47\cdot 10^{-6}\,s < \tau^c\, (1.89\, eV) $$
which translates (\ref{berg}) into
\begin{equation} \label{bmin}
b_{\min} = 1.35\cdot 10^{11} < b\, .
\end{equation}
That the constant $b$ has to be large is no surprise, as the action 
$b\hbar$ marks the transition from quantum to classical physics. The 
bound (\ref{bmin}) can easily be improved by estimating conventional 
effects which contribute to diminishing the contrast $K$. Beyond,
a direct measurement of a non-zero $\tau^c$ 
requires that all other effects can convincingly be controlled and that 
still a gap between the estimated and measured contrast remains. Such
an  analysis goes beyond the scope of the present paper. Here, I am
content with establishing firm, but crude, bounds on $b$.

Finally, in this paper, I derive an upper bound on $b$. Avalanche
photodiodes are the up-to-date devices for achieving
measurements in short time intervals, as needed for spacelike 
measurements \cite{Aspect}. Time resolutions down to $20\,ps$ FWHM
are achieved, see \cite{Co96} for a recent review. The energy 
consumption is sharply peaked in these short intervals
(order of watts), but does not translate into an immediate estimate 
of the collapse time. The reason is that collapse at some later time
may lead to indistinguishable results. Claiming differently includes 
the task of disproving popular decoherence ideas \cite{Zurek}.
Nevertheless, there is an easy way to estimate upper bounds by
analysis of actually working measurement devices: Our approach makes
only sense when the reduction process does keep up with the 
{\it sustained} performance of every real, existing measurement device.
Then, the energy dissipation of such a device yields immediately
an upper bound on $b$. Ref.\cite{Co96} gives on p.1964 the example
of a photo avalanche diode which operates at $10^5\,cps$ and has
a mean power dissipation of $4\,mW$. This translates into an energy 
consumption of about $2.5\cdot 10^{11}\,eV$ per count, {\it i.e.}
$$ \tau^c\, (2.5\cdot 10^{11}\, eV) <
\tau^c_{\max}\,(2.5\cdot 10^{11}\, eV) = 10^{-5}\, s\, ,$$
which implies
\begin{equation} \label{bmax}
b < b_{\max} = 3.8\cdot 10^{21}\, .
\end{equation}
Equation (\ref{bmin}) and (\ref{bmax}) leave a wide range open. An
analysis of existing experiments should allow to narrow things
down by a least a few orders of magnitude. Here the emphasize is
on quoting save, instead of sophisticated, bounds. Even this has
caused some efforts, the reason simply being that experimentalists
do not focus on the information needed. 

In conclusion, we have discussed the possibility of attributing to
microscopic matter the ability to perform wave function reductions.
It is of interest to improve the bounds $b_{\min}$ (\ref{bmin}) and
$b_{\max}$ (\ref{bmax}) for the collapse time $\tau^c (\triangle E)$ of
equation~(\ref{berg}). From this viewpoint, I would
like to argue in favor of a paradigm shift concerning QT experiments. 
It is no longer of central interest to demonstrate the existence of 
one or another exotic interference effect. We know, they are there. 
Most interesting is to control that interference happens for every 
single, participating particle. This puts the focus on experiments 
with high visibility. If one could convincingly demonstrate that 
particles occasionally skip participation in an interference pattern,
such a results could pave a major inroad towards understanding of the 
measurement process. The aim of pushing experiments towards optimal 
visibility is of interest in itself. Independent of its validity, 
the introduced collapse scenario provides an interesting classification 
pattern for such results: The achieved lower bounds $b_{\min}$ should
be compiled. Concerning $b_{\max}$, one is lead to minimizing the energy
dissipation of measurement devices under sustained performance. Again, 
this is a goal of interest in itself. 
\medskip

\noindent
{\bf Acknowledgements:} I would like to thank Dr. Wolfgang Beirl for
his interest and useful discussions.

\section*{Figure Captions}

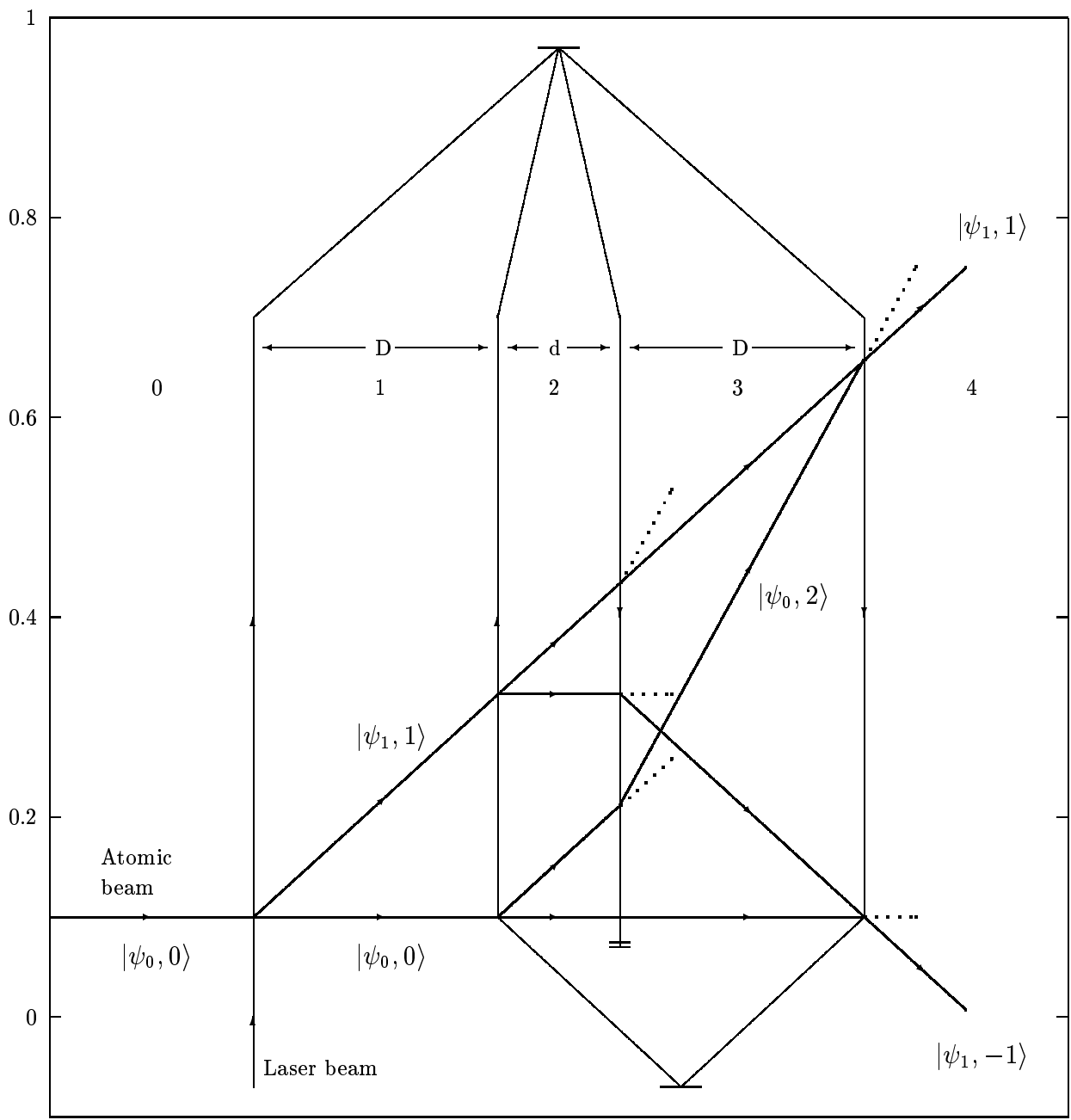
\noindent{\bf Figure 1:} Bord\'e interaction geometry \cite{Bo84,St97}
of four traveling laser beams to 
create optical Ramsey fringes in atomic spectroscopy. The atomic beam
is incoming from the left and (split in partial waves) outgoing to the
right. Interactions zones are where the atomic beam crosses the vertical
lines of the traveling laser beam. The numbers $i=1,2$ and $3$ label
free propagation ranges between the interaction zones. The numbers $0$
and $4$ label the free propagation ranges before the first and after the
last interaction.

\end{document}